\documentclass[]{aa}
\usepackage{times}
\usepackage{graphicx}
\newcommand{\Lx}{L_{\rm X}}
\newcommand{\LT}{L_{\rm X}-T_{\rm X}}
\newcommand{\Tx}{T_{\rm X}}
\newcommand{\Fx}{F_{\rm X}}

\begin{document}

%   \thesaurus{12.03.1;12.03.3;12.03.4;12.04.2;12.12.1;11.03.1}  

\title{Introducing BAX: a database for X-ray clusters and groups of galaxies}

\titlerunning{BAX X-ray cluster database}
  
\author{
 R.~Sadat
\inst{1},
 A.~Blanchard
\inst{1},
M. ~Arnaud
\inst{2},
 J.-P. ~Kneib
\inst{1,4},  
G. ~Mathez
\inst{1},
B. ~Madore
\inst{3} 
J.M. ~Mazzarella
\inst{3}
          }
 
  \institute{
%%1
Laboratoire d'Astrophysique de l'Observatoire Midi-Pyr\'en\'ees, 14 Avenue. E. ~Belin, F--31400 Toulouse, France
\and
%%2
CEA/DSM/DAPNIA Saclay, Service d'Astrophysique, l'Orme les Merisiers B\^at 709, 91191 Gif-sur-Yvette, France
\and
%%3
California Institute of Technology, Jet Propulsion Laboratory, MS 100-22, Pasadena, CA 91125
\and
Astronomy Department, California Institute of Technology, Pasadena, CA 91125
%%5
}

 \offprints{R. Sadat, {\tt rsadat@ast.obs-mip.fr}}

%\date{\today}
\date{Received \rule{2.0cm}{0.01cm} ; accepted \rule{2.0cm}{0.01cm} }
\authorrunning{Sadat \& al. }

%________________________________________________________________

\abstract{
We present BAX, Base de Donn\'ees Amas de Galaxies X 
(http://webast.ast.obs-mip.fr/bax), a multi-wavelength database dedicated to 
X-ray clusters and groups of galaxies allowing detailed information retrieval. 
BAX is designed to support astronomical research by providing 
access 
%the astronomers with 
to published measurements of the main physical quantities 
%in the X-Ray wavelength 
and to the related bibliographic references:
basic data stored in the database are cluster/group identifiers, equatorial
coordinates, redshift, flux, X-ray luminosity (in the ROSAT band) and
temperature, and links to additional linked parameters 
(in X-rays, such as spatial profile
parameters, as well as SZ parameters of the hot gas, lensing measurements,
and data at other 
wavelengths, such as optical and radio).
The clusters and groups 
in BAX can be queried by the basic parameters as well as the
linked parameters or combinations of these.
We expect BAX to become an important tool for 
the astronomical community. BAX
will optimize various aspects of the scientific analysis of X-ray clusters
and groups of galaxies,
from proposal planning to data collection, interpretation and publication,
from both ground based 
facilities like MEGACAM (CFHT), VIRMOS (VLT) and space missions like
XMM-Newton,  
Chandra and Planck.  
}

%________________________________________________________________
\maketitle

\section{Introduction}
Clusters of galaxies are the largest virialized systems in the Universe and
hence are thought to be useful probes of the universe. 
Historically, they have
played an important role in cosmology, as they provided the first evidence
 of the presence of substantial amounts of dark matter in the Universe 
(Zwicky 1933). 
During the 70's, it has been discovered that galaxy clusters are strong X-ray 
emitters and it has been realized that crucial information can be obtained 
from the X-ray data and several space missions have been devoted to X-ray 
observations of clusters of galaxies (HEAO-1,2, Ariel V, Einstein, EXOSAT, 
Ginga, Tenma, ROSAT, ASCA) and more recently XMM-Newton and Chandra.
X-ray observations of clusters of galaxies have increased rapidly in the last 
years with the advent of good quality imaging and spectroscopy. All these 
observations resulted in large amount of X-ray data for thousands groups and 
clusters of galaxies published in an exponentially increasing number of papers.

The study of the statistical properties of X-ray clusters and groups 
and their evolution 
with redshift (i.e. their temperature and luminosity functions, as well as
their luminosity-temperature relation) allow  to constrain the 
cosmological parameters and to place strong constraints on large scale 
structure theories (Peebles et al. 1989; Oukbir \& Blanchard 1992; Hattori
\& Matsuzawa 1995;
Eke et al. 1996; see Rosati et al. 2002 for a review). 
The X-ray properties of groups and clusters indicate important
effects 
of non-gravitational processes (Kaiser 1991; Evrard \& Henry 1991), which
have fundamental consequences on the theory of galaxy formation.
 The study of chemical abundances in the intracluster cluster
 medium and its evolution should provide us with valuable information on metal
 enrichment of the universe and the history of nucleosynthesis in the universe
 and therefore on the star formation history (Mushotzky \& Lowenstein 1997). 

Moreover clusters of galaxies are extended and often complex 
objects that can be observed at 
several wavelengths, 
% the number of physical quantities relevant to their 
% description leads 
% {\bf mentionner XMM et Chandra}
leading to an increasing amount of scientific data and publications.
% Consequently,
However, access to this information through World Wide Web (WWW) 
is  limited. Although some information (catalogs, tables, 
bibliographic references...) can be retrieved from existing databases such as 
NED (Mazzarella et al., 2001), SIMBAD (Wenger et al., 2000) or VIZIER 
(Ochsenbein, Bauer and Marcout, 2000), more specific information on X-ray 
clusters properties cannot presently be easily obtained.   

We attempt to provide an easy X-ray cluster parameter retrieval, by
introducing the BAX (Base de donn\'ees d'Amas de galaxies X) X-ray cluster
online database, which is designed to support scientists, space 
missions and ground based observatories in the planning, interpretation and 
publication of research on galaxy clusters. 
% BAX provides access 
% to the growing number of data on X-ray galaxy clusters that can be searched 
% by name, positions, redshift, flux, luminosity, temperature and
% bibliographical 
%  keywords using a Web user interface accessible from any Web browser. 
Such a database is strongly needed and should be particularly 
useful to
%appreciated by 
the astronomical community for the analysis and understanding of the global 
properties of groups and clusters and the correlations between their
physical properties. 

\section{Main objectives}

The primary goal of BAX is to help scientists working on X-ray clusters and
groups,
in their activity 
%working 
from proposal planning 
to data collection, reduction, interpretation and publication, 
by allowing a rapid and efficient access to the ensemble of 
published data measurements and bibliographical references on existing X-ray
 galaxy clusters and groups,
through a comprehensive and an easy-to-use tool. In its first
implementation, BAX therefore:
\begin{itemize}
\item allows access to all published {\it basic data} and the
corresponding bibliographic references;
\item allows data collection and sampling through various selection criteria;
\item helps in preparing proposals for future X-ray observations.
\end{itemize}
In a second stage, BAX will provide users with new capabilities that 
will be defined on the basis of their needs.
These future improvements are described in section~\ref{improve}. 

\section{How does BAX operate?}
For a given cluster or group, BAX contains:
\begin{itemize}
\item the J2000 equatorial coordinates ($\alpha$, $\delta$) and redshifts
$z$,  generated automatically from the NASA/IPAC Extragalactic Database (NED);
\item a set of basic X-ray measurements:  X-ray fluxes ($\Fx$) in ROSAT band,
 X-ray
luminosities ($\Lx$) converted in the ROSAT [0.1--2.4 keV] band, and the X-ray
temperatures ($\Tx$); 
\item the corresponding bibliographical references.
\end{itemize}
Depending on the menu query, the user can search for an individual cluster by
name or for an ensemble of clusters that meet chosen selection criteria, such
as equatorial coordinates, redshifts, cluster basic data ($\Fx$,$\Lx$,$\Tx$)
or on linked parameters (see the help page
{\tt http://bax.ast.obs-mip.fr/bax-help.html} for the list of such linked
parameters).   

\subsection{Query functions through the Web interface}
BAX is accessible through the WWW interface located at 
{\tt http://bax.ast.obs-mip.fr} site. The user starts his search 
through the main menu. By now five query modes are 
proposed: {\sf By name}, {\sf By positions}, {\sf By parameters}, {\sf By
keywords}  
and  {\sf Multi-criteria}, this latter  menu allows the combination of  all 
the modes. BAX will then retrieve data and/or  bibliographic references on a
given cluster or on a list of clusters that respond to the chosen
criteria. The outputs are:
\begin{itemize}
\item the cluster acronyms;
\item the equatorial J2000 coordinates ($\alpha$, $\delta$)
\item the cluster redshift $z$;
\item the {\it canonical measurement} of the three basic data $\Fx$, $\Lx$
and $\Tx$.
\end{itemize}
The {\it canonical measurements} are selected among recent and
accurate existing measurements, and $\Fx$ and $\Lx$ are homogenized
by conversion to the ROSAT [0.1--2.4 keV] energy band.  
The user can ask for further information to get access to all published measurements of the basic data through the {\sf all measurements} query and/or access to an ensemble of bibliographic references selected by keywords criteria through the {\sf Bibliography} criteria. The bibliography search will retrieve the papers associated to at least one of the chosen keywords.

%\section{Tools}

The heart of BAX is the database with its Web interface query service, but BAX
 is also conceived to provide the user with various tools to 
ease and optimize the use of the database as well as to allow data analysis. 
For instance, BAX contains the coordinate conversion and precession 
and the flux converter that allows to convert fluxes from any given energy 
band to the [0.1-2.4 keV] ROSAT reference band.

\subsection{Establishing the $Temperature-Luminosity$ relation using the ``By multi-criteria'' query}

%\begin{figure}[h]
%\includegraphics[scale=0.5,angle=0]{bymulticrit_L_T.ps}
%\caption {Exemple of ``Multi-criteria'' query: clusters with available luminosity {\bf and} temperature measurements}
%\label{fig:fig_multicrit}
%\end{figure}
%\begin{figure}[h]
%\includegraphics[scale=0.5,angle=0]{bymulticrit_L-t-result.ps}
%\caption {The result of {\sf Multi-criteria} query. BAX found 247 clusters with available temperature and luminosity measurements.}
%\label{fig:fig_multicrit_result}
%\end{figure}
\begin{figure}[h]
\includegraphics[scale=0.5,angle=0]{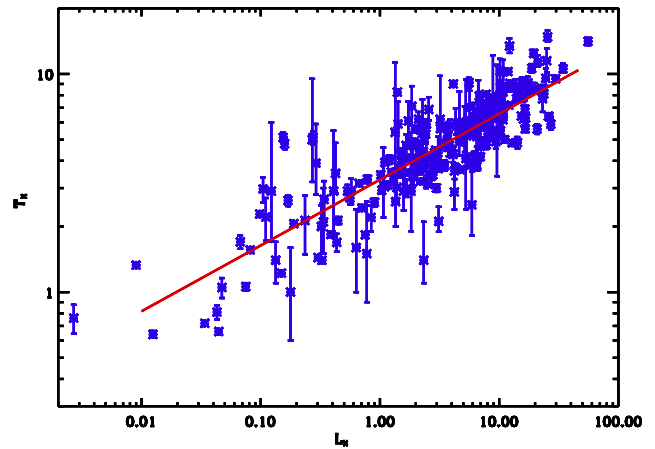}
\caption{ The $\LT$ relation as obtained using the 247 clusters with measured $\Lx$ and $\Tx$ found in BAX database. The (red) solid line is the result of fitting $\Tx$ versus $\Lx$ to the data. We find a correlation slope of 3.3. Note here that luminosities we use are luminosities in the ROSAT [0.1-2.4] rest-frame band.}
\label{fig:LT}
\end{figure}
In the following we will illustrate an exemple of BAX capabilities through the ``Multi-criteria'' query. As already mentionned, this menu allows a more elaborated query by 
combining criteria on name, positions, physical parameters and keywords. With this feature one can easily build up a sample of galaxy clusters selected on
 a set of criteria, (useful for statistical studies).
An example of query is to build a sample of clusters with available luminosity and  temperature measurements, in order to study one of the best studied correlations among X-ray clusters the $Luminosity-Temperature$ relation (Mitchell et al. 1979, Edge $\&$ Stewart 1991, David et al. 1993, Arnaud $\&$ Evrard 1999). This relation between the X-ray temperature $\Tx$ and luminosity $\Lx$ is important because it is indicative of the internal structure in clusters of galaxies and possible variations in the gas mass fractions (the ratio of the X-ray gas to total mass) since the luminosity is determined by the mass of the intra-cluster medium emitting the hot X-ray gas and the temperature is determined by the total gravitating mass. The user can retrieve such a sample by simply filling the 
adequate fields in the multi-criteria menu form (here the $\Lx$ and $\Tx$ fields). BAX will return a list of clusters that meet all the 
criteria. For each cluster of the sample BAX provides with the basic data $\alpha$, 
$\delta$, $z$, $\Fx$, $\Lx$ and $\Tx$ 
%(see Fig. (\ref{fig:fig_multicrit_result}). 
Here again further information can be obtained for each single cluster of the list through the {\sf all 
published measurements} and/or the {\sf bibliography}. Figure 1 presents the resulting $\LT$ relation we derived from the luminosity and temperature  measurements we found using BAX multi-criteria query.

\section{Current status}

The number of published papers on X-ray clusters and groups 
is increasing rapidly, and
this rise 
should hold in the forthcoming years because of existing and
future space and ground facilities. 
Updating the database is therefore a long-term task that is necessary for its 
scientific relevance. By now, the most important large X-ray cluster
catalogs have been included in the database, except for catalogs that are 
not yet published such as the MAssive Cluster Survey (MACS, Ebeling et
al. 2001), the ROSAT-ESO flux limited X-ray (REFLEX) cluster survey
(B\"ohringer et al. 2000) and the ROSAT Deep Cluster Survey 
(RDCS, Rosati et al. 1995, 1998). BAX is currently kept up-to-date on a daily 
basis by integrating the most recent bibliographic references mainly resulting 
from Chandra and XMM-Newton observations very soon after their publication in
journals referenced in ADS. At the time of writing, 
BAX is complete in terms of XMM-Newton publications and published papers during year 2003 and it contains data on  1579 clusters and groups, 
with 8116 measurements and 33185 linked 
keywords, among which 
1371 clusters and groups with available flux measurements in BAX and 298 with
temperature 
measurements. The published version of the database was released in
September 2002. The ``debugged'' version of the database is running since May
2003. 7949 connections and 3938 queries have been recorded since that
date. During the year 2003, 
%the statistics are going increasing and
BAX recorded roughly
300--400 connections per month with approximately the same number of
queries, and these rates are increasing in 2004.

\section{BAX updates and future improvements}
\label{improve}

BAX has been conceived to be adaptable: 
for example, the list of linked keywords can be easily
modified by the administrator and linked parameters
can be turned into basic parameters stored in the database.
Beyond existing data on galaxy clusters and groups, BAX will evolve
 to offer new 
functionalities accordingly to the needs of users.
For instance, among various requests that were asked for are:
\begin{itemize}
\item the direct access to other existing X-ray archives;
\item the access to properties of virtual clusters obtained from numerical
simulations;
\item the addition of dynamical links to other data services;
\item the extension to non X-ray galaxy clusters and groups: optical, and
 radio parameters, as well as
 gravitational 
 lensing and SZ parameters;
\item  the implementation of new tools such as: XMM-Newton data analysis
using the SAS software, tools to visualize the query outputs (histograms,
charts);
\item the development of a portal for public outreach.
\end{itemize}
The extension to non X-ray observations is motivated by several large scale
facilities, either ongoing or scheduled in the future, 
which will provide us with massive  
sets of data concerning known clusters and groups,
as well as the  discovery of numerous 
new such systems: large scale optical surveys (e.g. SDSS, CFHT Legacy Survey,
GOODS),  will produce new catalogs of optically selected  clusters, at the same time weak 
lensing surveys from ground-based (e.g. CFHT Legacy Survey) 
or space missions are going to provide direct 
lensing detection of dark matter concentrations, perhaps pointing to dark
groups and clusters (Erben et al. 2000). 
In the near future, the Planck Surveyor and ground-based telescopes such as 
APEX, SPT, ACT should produce large catalogs of SZ detected clusters (Barbosa
et al, 1996; Carlstrom et al. 2001). In this context, BAX 
will be extremely useful at several levels: in its present version, it will
allow to obtain the cross identification with known X-ray clusters, but also
it will ease the statistical correlation of the survey clusters according to 
various physical properties. Finally, BAX will be one of the 
privileged ways to facilitate the distribution of survey results. 

At a more sophisticated level, BAX 
might provide the user with  tools for quick look  data product 
from  ongoing missions. It is also  
intended to plug as an on-line web server tool an automated pipeline
(described in Marty 2003) to reduce 
EPIC MOS \& PN spectro-imaging data. 
%This pipeline is described in Marty 2003.
 More general software tools for X-ray data analysis, simulation and proposal 
preparation tools will also be available as a service provided by BAX, by 
using public on-line version of existing tools 
such as WebPIMMS, XANADU, FTOOLS and
other HEASARC (High Energy Astrophysics Science Archive Research Center) 
software tools.   

\section{BAX in the context of a global virtual observatory}

Among the purposes of the {\it Virtual Observatory} (VO) is the
interconnection and 
federation of  
data archives, surveys and software tools from observatories using 
common database query standards and data transfer protocols. The need of such a
VO is driven by the growing size of astronomical data sets due to new large 
facilities and the ambition to discover new scientific results 
from connected large multi-wavelength data sets using data mining tools. In 
this context, BAX is able to provide its specific contribution to the 
construction of the global virtual observatory, including the bibliographic 
aspect. We expect that the future 
developments, such as the extension to  multi-wavelength published data on 
galaxy clusters and groups, the implementation of on-line data analysis and
simulation software tools and the
interoperability with other archival services will allow to specify in  more 
details  the role that BAX could play in the Virtual Observatory.

\section{Brief summary}
BAX is an online database of X-ray clusters and groups of galaxies designed to 
support scientists at different steps of their research from proposal planning 
to data collection, interpretation and publication. 
BAX provides published data measurements of the basic properties of
 clusters and groups, including pointers to related bibliographical
references. The ultimate 
 goal of BAX is to provide not only an interface to query its database but also
to serve as a portal for the general community of scientists working on
clusters and groups of galaxies. We expect BAX
 to become an essential tool for the astronomical community especially in view 
of preparing missions like Planck Surveyor, which will detect up to 50,000
individual  
galaxy clusters through the SZ effect on the cosmic microwave 
background (Kay et al. 2001). 
In the future, we hope to develop new capabilities and services by developing 
new collaborations with other organizations such as the CDS (Strasbourg), NED 
(Caltech), HEASARC (NASA), LEDAS (Leicester), as well as with the
XMM-Newton  and Chandra X-Ray observatories, in view of contributing to the
construction of the {\it Virtual Observatory}.

\section{Acknowledgments}
BAX has been supported by the Centre National d'Etudes Spatiales (CNES), 
the french Programme National de Cosmologie (PNC) and the Observatoire de 
Midi-Pyr\'en\'ees (OMP). We thank S. Drake and the HEASARC team for the
announcement on BAX in the HEASARC Home Page. We also would like to thank
G. Mamon for reading the paper and for useful discussions.

\end{document}